\documentclass[twocolumn,showpacs,english,prb,superscriptaddress,preprintnumbers,amsmath,amssymb,floatfix]{revtex4}

\usepackage[T1]{fontenc}
\usepackage[latin9]{inputenc}
\usepackage{epsfig,psfrag}
\usepackage{dcolumn}
\usepackage{bm}
\usepackage{graphicx}
\usepackage{color}
\usepackage{esint}
\usepackage{babel}

\begin{document}

\title{Coulomb blockade of Majorana fermion induced transport}

\author{A. Zazunov}

\affiliation{ Institut f\"ur Theoretische Physik, 
Heinrich-Heine-Universit\"at, D-40225  D\"usseldorf, Germany }
 
\author{A. Levy Yeyati}

\affiliation{Departamento de F{\'i}sica Te{\'o}rica de la 
Materia Condensada C-V,
Universidad Aut{\'o}noma de Madrid, E-28049 Madrid, Spain }

\author{R. Egger}

\affiliation{ Institut f\"ur Theoretische Physik, 
Heinrich-Heine-Universit\"at, D-40225  D\"usseldorf, Germany }

\date{\today}

\begin{abstract}
We study Coulomb charging effects for transport through a
topologically nontrivial superconducting wire, where Majorana bound states
are present at the interface to normal-conducting leads. 
We construct the general Keldysh functional integral representation,
and provide detailed results for the nonlinear current-voltage
relation under weak Coulomb blockade conditions.
\end{abstract}

\maketitle

\section{Introduction}
\label{sec1}

Charge transport through topologically nontrivial materials is of 
great current interest, offering novel fundamental insights
as well as potential applications in topological 
quantum computing.\cite{nayak} The recently discovered topological insulators 
as well as topological superconductors (TSs)\cite{hasan,qizhang} 
are predicted to exhibit spectacular nonlocal transport phenomena
and have consequently attracted a lot of attention.
In particular, for one-dimensional TS quantum wires,
the crucial role of Majorana bound states (MBSs) located near the interface to 
topologically trivial regions has been 
emphasized.\cite{kitaev,fukane,tanaka,law,linder}
Majorana fermions are special in that they are their own antiparticles,
i.e., the fermion creation operator is equal to the annihilation operator,
and the condensed-matter setup discussed below could offer 
experimental signatures for these elusive and hitherto unobserved particles.

\begin{figure}
\centering
\includegraphics[width=8cm]{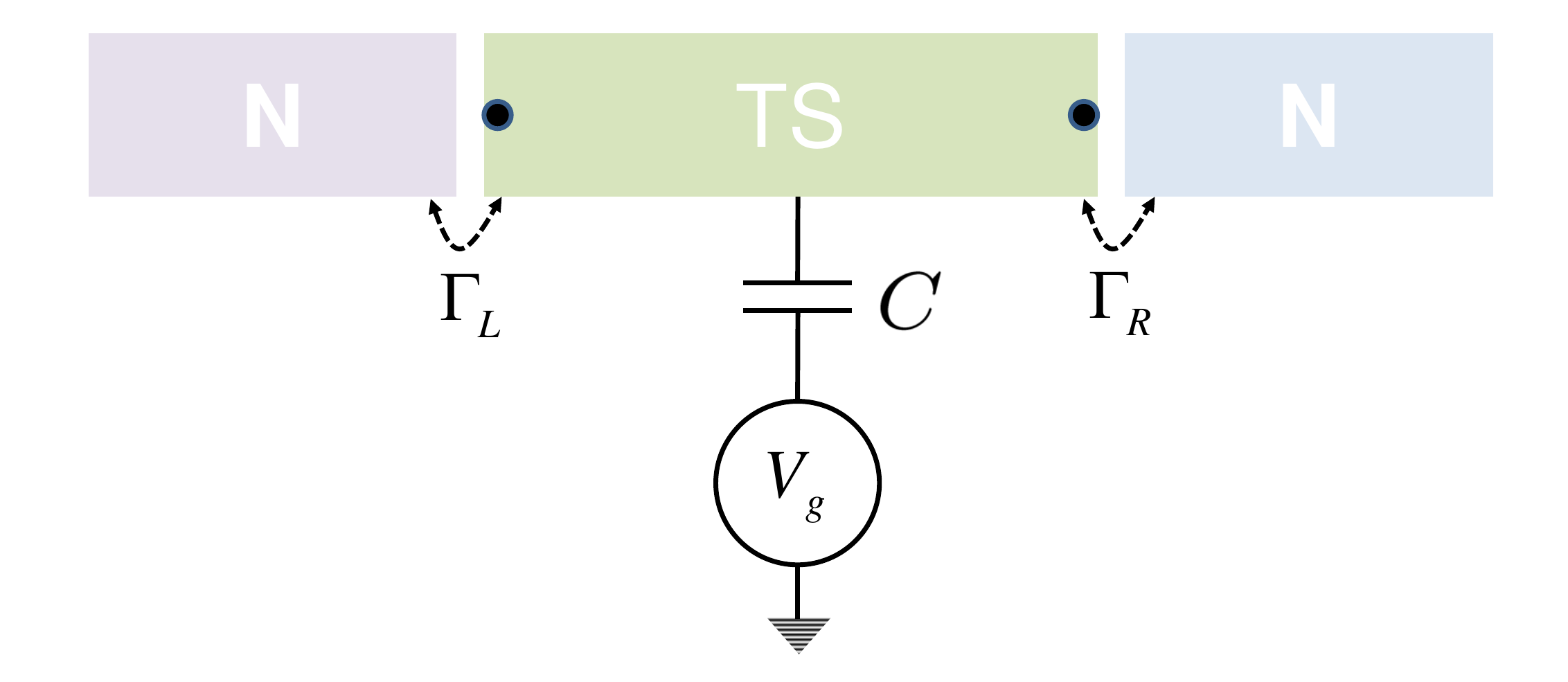}
\caption{\label{fig1} (Color online) Setup studied in this work: a topological
superconductor (TS) dot containing Majorana bound states at 
the ends (indicated by black dots) is coupled via tunnel contacts 
(hybridizations $\Gamma_{L,R}$) to normal-conducting (N) 
metallic electrodes.  The charging energy $E_c=e^2/(2C)$ causes phase 
fluctuations on the TS dot and influences the current-voltage relation.
We assume that the superconducting gap $\Delta$ of the TS is large
compared to the other energy scales, i.e., quasiparticle excitations
on the dot can be neglected. }
\end{figure}

We here study nonlinear electronic transport through a mesoscopic
TS wire (also referred to as ``dot'') containing a pair of MBSs.
The wire is assumed to be contacted by conventional normal-conducting leads.  
The contacts correspond to thin insulating layers or 
barriers and will be modeled by a tunnel Hamiltonian.  
We include Coulomb interactions on the dot 
via the charging energy $E_c=e^2/(2C)$.  The schematic setup
studied in this work is shown in Fig.~\ref{fig1}, and 
we briefly outline possible experimental realizations below.
The noninteracting  ($E_c=0$) version of this setup was
studied in detail before,\cite{law,bolech,nilsson,flensberg,golub} 
and one finds \textit{resonant Andreev reflection} with the maximal 
(unitary) value $G_{L/R}=2e^2/h$ of the linear conductance 
 through the left/right contact.
For the interacting case ($E_c>0$), an important work by 
Fu\cite{fu} has provided a general framework to address 
transport in this setup.  The Hilbert space constraint
introduced in Ref.~\onlinecite{fu} is, however, difficult to handle
in actual calculations. In fact, the concrete results presented
in Ref.~\onlinecite{fu} (see also below) were mostly restricted
to the strong Coulomb blockade (CB) regime realized for large $E_c$. 
We formulate an alternative Keldysh functional integral
approach below, where the constraint is automatically satisfied.
We give detailed results for the weak CB  (small $E_c$) regime, where
we find a pronounced interaction ($E_c\ne 0$) induced suppression of 
the current.

Experimental realizations for the setup shown in Fig.~\ref{fig1} arise
under several distinct physical scenarios. For instance,
a nanowire made of a 3D topological insulator (e.g., Bi$_2$Se$_3$) 
with proximity-induced superconducting correlations is predicted to have
MBSs at both ends.\cite{franz} 
Similar setups can be realized in 2D topological insulators with 
``quantum spin Hall'' edge states.\cite{fukaneqsh,fu}  Another example
is based on the vortex core states in $p$-wave superconductors\cite{bolech} 
or in 2D noncentrosymmetric superconductors.\cite{sato}  Finally,
the MBS should also be realizable in semiconductor quantum wires
with a Zeeman field and proximity-induced superconductivity.\cite{felix}
The stability of the MBS when including interactions in an isolated 
wire has also been addressed in several recent works.\cite{intwire}
However, so far no experimental signature for the MBS has been reported
in any system.  

The structure of this paper is as follows. In Sec.~\ref{sec2}
we describe the model employed in our study and briefly
show that our formulation reproduces known results\cite{fu}
in the strong CB limit.  We then construct 
a  general Keldysh functional integral representation for the interacting
problem in Sec.~\ref{sec3}. This theory is employed in Sec.~\ref{sec4}
to evaluate the current-voltage characteristics under weak CB conditions.  
We briefly summarize our main findings and provide an outlook in 
Sec.~\ref{sec5}.  Technical details and derivations can be found 
in three appendices.  Note that we often use units where $\hbar=k_B=1$.

\section{Model}
\label{sec2}

We consider transport through a mesoscopic TS quantum wire
contacted by normal-conducting leads, see Fig.~\ref{fig1}.
We assume throughout this work that 
the superconducting gap $\Delta$ is the largest relevant
energy scale, and thus no quasiparticles  above the gap are involved.
In particular, ${\rm max}(eV,E_c)<\Delta$.  We study a ``floating''
(not grounded) TS dot, where current must be conserved  
under steady-state conditions.  
In some previous works, see, e.g., Ref.~\onlinecite{bolech},
a grounded superconductor was considered, where current need not 
be conserved.   Note also that the situation is different
for conventional (topologically trivial) 
superconducting dots, where transport is typically mediated
by cotunneling processes and the resulting conductance is
much smaller.\cite{nazarov} In the present case,
two MBSs may host a fermion at no 
energy cost, and thus no even-odd parity effect\cite{nazarov} arises.

\subsection{Model}

For realistical TS wire length, the two MBSs located near the left/right 
($j=L/R$) contact correspond to
\textit{decoupled}\ Majorana fermion operators, $\gamma_j=\gamma_j^\dagger$,
with anticommutator relation $\{\gamma_i,\gamma_{j}\}=\delta_{ij}$. 
In the low-energy sector of interest here, only the MBSs
provide available fermion states to tunnel through the superconductor. 
It is convenient to introduce the complex auxiliary fermion 
$d= (\gamma_L+i\gamma_R)/\sqrt{2}$, i.e.,
\begin{equation}\label{aux}
\gamma_L= (d+d^\dagger)/\sqrt{2},\quad
\gamma_R= -i(d-d^\dagger)/\sqrt{2}.
\end{equation}
For an isolated wire, the fermion parity $(-)^n$ of the two degenerate
ground states with even $(n=2N)$ or odd ($n=2N+1$) number of electrons 
fixes the occupation of the $d$ fermion level,
\begin{equation}\label{aux2}
2i\gamma_L\gamma_R = 2 \hat n_d  -1 = (-)^{n+1}, \quad \hat n_d=d^\dagger d.
\end{equation}
This imposes a constraint on the Hilbert space,\cite{fu} 
and the Majorana  fermion dynamics cannot be taken as independent 
of the charge (or the dual phase) dynamics. Below, we instead formulate 
an alternative but essentially equivalent approach
free of any Hilbert space constraint, see also Ref.~\onlinecite{vanheck},
which is technically easier to handle.
To that end, consider the number operator $\hat N$ for Cooper pairs
conjugate to the condensate phase $\chi$ on the dot,
$[\chi,\hat N ] = i$.  The dot's instantaneous charge state is then 
fully determined by specifying the configuration $(N,n_d)$, 
where $N\in \mathbb{N}_0$ and $n_d=0,1$ are eigenvalues of 
$\hat N$ and $\hat n_d$, respectively.
Importantly, Eq.~\eqref{aux2} tells us that the fermion
parity of the dot is determined by $n_d$ only and not affected
by changes of $N$. By construction, both $\hat N$ and $\chi$ commute
with $\gamma_j$, and therefore no constraint on the Hilbert space arises.

The full Hamiltonian describing transport through the dot,
$H=H_c+H_t+H_l$, contains the Coulomb charging term
\begin{equation}\label{hc}
H_c = E_c (2\hat N+ \hat n_d -n_0)^2,
\end{equation}
where $n_0\in \mathbb{R}$ is tunable by the backgate voltage $V_g$ indicated
in Fig.~\ref{fig1}.  Electrons in the leads
correspond to effectively spinless fermion operators $c_{j=L/R,k}$ 
with momentum $k$, where for given $j$,
$c_{jk}$ and $c^\dagger_{jk}$ are tunnel-coupled 
to the respective Majorana fermion $\gamma_j$ only. 
For simplicity, we assume a $k$-independent tunnel matrix element
 $\lambda_j$ encoding the overlap  of the lead state with the
respective MBS wavefunction.  As detailed in App.~\ref{appa},
the tunnel Hamiltonian then takes the form
\begin{eqnarray}\label{ht1}
H_t &=& \frac{1}{\sqrt{2}} \sum_{k} \Bigl[ \lambda_L  c_{L,k}^\dagger 
(d+e^{-i\chi} d^\dagger) \\ \nonumber
&-& i \lambda_R c_{R,k}^\dagger (d-e^{-i\chi}d^\dagger) \Bigr]
 + {\rm h.c.},
\end{eqnarray}
which is charge conserving. In particular, the different 
terms in $H_t$ describe electron tunneling from the dot to lead $j$, either 
by destroying the $d$ state without changing $N$, i.e.,
$(N, 1)\to (N,0)$, or by occupying the $d$ state together with a 
splitting of a Cooper pair, i.e., $(N,0)\to (N-1,1)$, plus the 
conjugate processes.  As elaborated in Appendix \ref{appb}, 
we treat the lead Hamiltonian $H_l$ 
within the standard wide-band approximation.\cite{naz,altl}  
The applied bias voltage $V$ corresponds to 
the chemical potential difference in both leads, $eV= \mu_L-\mu_R$.
For further reference, we also define the hybridization energy scales
\begin{equation}\label{hyb}
\Gamma_j = 2\pi \nu_j |\lambda_j|^2,
\end{equation}
where $\nu_j$ is the density of states in lead $j=L/R$. 

It is now straightforward to rederive the well-known 
results\cite{bolech,nilsson,flensberg,golub} for the
noninteracting ($E_c=0$) limit from the above Hamiltonian,
see also Sec.~\ref{sec4}.  
Before turning to the construction of the Keldysh functional
integral for the interacting problem in Sec.~\ref{sec3},
let us briefly show that the 
above formulation easily reproduces previous results\cite{fu} reporting
electron ``teleportation'' under strong CB
conditions, i.e., for ${\rm max}(eV,\Gamma_{L,R})\ll E_c$. 

\subsection{Strong Coulomb Blockade}

Under strong CB conditions, the charging 
energy (\ref{hc}) is dominant and needs to be examined first.
We focus on the resonant situation realized for half-integer values 
for $n_0$, where two different cases arise:
(i) For $n_0=2\ell +1/2$ with integer $\ell$, 
degeneracy is achieved for fixed Cooper pair number $N=\ell$,
where the $d$ state is either occupied ($n_d=1$) or empty ($n_d=0$).
(ii) When $n_0=2\ell -1/2$, the degeneracy point is achieved
for different Cooper pair numbers, namely for 
$(N,n_d)=( \ell-1, 1)$ and $(\ell,0)$. 
We conclude that the low-energy description needs to keep
only two states, either of type (i) or (ii) depending on $n_0$.
For case (i), $N$ remains constant and
all terms $\propto e^{\pm i\chi}$ (where $N$ is raised or lowered 
by one unit) can be omitted in Eq.~\eqref{ht1}.  The tunnel
Hamiltonian then takes the form
\[
H^{(i)}_t = \frac{1}{\sqrt{2}} \sum_{k} 
\left( \lambda_L  c_{L,k}^\dagger  - i \lambda_R c_{R,k}^\dagger\right) d
+ {\rm h.c.}
\]
Under scenario (ii) only transitions $(\ell-1,1)\leftrightarrow (\ell,0)$
are possible.  Using $f=e^{-i\chi} d^\dagger$ as effective single-charge
fermion, we arrive at 
\[
H^{(ii)}_t = \frac{1}{\sqrt{2}} \sum_{k} 
\left( \lambda_L  c_{L,k}^\dagger  + i \lambda_R c_{R,k}^\dagger\right) f
+ {\rm h.c.}
\]
In both cases, we recover the resonant tunneling model
describing teleportation.\cite{fu} This leads to 
the unitary conductance value $G_L=G_R=e^2/h$ instead 
of the resonant Andreev reflection value $2e^2/h$ found when $E_c=0$.

\section{Keldysh functional integral}
\label{sec3}

We now turn to the general Keldysh functional
integral\cite{naz,altl} for the interacting problem.  This
corresponds to the Majorana generalization of the (real-time version of the) 
Ambegaokar-Eckern-Sch\"on (AES) action for a conventional 
metallic grain.\cite{zaikin,dot,altland}  

\subsection{AES action for Majorana induced transport}
\label{sec3a}

The relevant low-energy degree of freedom characterizing 
the interacting problem is the single-electron (half Cooper pair) 
counting phase field $\phi(t)\equiv \chi/2$.    
For the nonequilibrium problem at hand, we employ the textbook Keldysh 
functional integral formulation\cite{naz,altl} and double 
the phase field $\phi(t)\to\phi_\pm(t)$ according
to the forward and backward branch of the Keldysh time contour.
Technically, the phase field is introduced through a 
Hubbard-Stratonovich transformation
of the charging energy in Eq.~\eqref{hc}, see~App.~\ref{appb} for details.
Physically, $\phi$ is the dual variable to the Cooper pair charge $N$.
We then introduce classical/quantum phase fields,\cite{dot,altland}
\begin{equation}\label{phicq}
\phi_c(t)= \frac12 (\phi_++\phi_-),\quad \phi_q(t)= \phi_+-\phi_-.
\end{equation} 
The charging energy (\ref{hc}) yields the action contribution
\begin{equation}\label{sc11}
S_c = \int dt \ \dot \phi_q \left( \frac{\dot\phi_c}{2E_c}+n_0 \right)
\end{equation}
encoding Coulomb blockade effects.  
By virtue of the Hubbard-Stratonovich transformation, 
the fields corresponding to the lead fermions and 
to the Majorana fermions effectively become noninteracting.
In the functional integral, those fields can therefore be integrated
out analytically; for technical details, see Appendix \ref{appb}.  
The final (phase representation of the) functional integral 
for the Keldysh partition function reads
\begin{equation}\label{keld}
{\cal Z} = \int {\cal D}(\phi_c ,\phi_q) \ e^{i (S_c+S_f)}.
\end{equation}
The fermion field integrations result in the action piece
\begin{equation}\label{sf11}
S_f = -i \sum_{j=L/R} \ln {\rm Pf} \left [ \check \tau_x i \partial_t 
\delta(t-t') + \check \Lambda_j(t,t') \right ],
\end{equation}
where ``Pf'' denotes the Pfaffian and 
$\check\tau_{x,y,z}$ are Pauli matrices in (rotated\cite{altl}) 
Keldysh space.  Moreover, we define the antisymmetric (time and Keldysh)
 matrix
\begin{widetext}
\begin{eqnarray} \label{antilamb}
\check\Lambda_j(t,t') &=&  2 i\Gamma_j F(t-t') 
\left( \begin{array}{cc} c(t) c(t') \cos\Phi_j(t,t')
& c(t) s(t') \sin\Phi_j(t,t') \\ 
- s(t) c(t') \sin\Phi_j(t,t') & s(t) s(t') \cos\Phi_j(t,t') \end{array}
\right) \\ \nonumber
& +& i\Gamma_j  \left(\begin{array}{cc}  0 & c^2(t) \delta_-(t-t')-s^2(t) 
\delta_+(t-t') \\ -c^2(t)\delta_+(t-t') +s^2(t)\delta_-(t-t') & 0 
\end{array}\right) 
\end{eqnarray}
\end{widetext}
with $c(t)=\cos[\phi_q(t)/2],  s(t) =\sin[\phi_q(t)/2]$, and
$\delta_\pm(t)\equiv \delta(t\pm 0^+)$, where the infinitesimal shifts 
reflect the proper causality features.  At temperature $T$, the 
distribution function $F(t-t')$ has the Fourier transform 
\begin{equation}\label{dist}
F(\epsilon)=\tanh\left[\epsilon/(2T)\right],
\end{equation}
and the phase function $\Phi_j$ appearing in Eq.~\eqref{antilamb} is
\begin{equation} \label{Phi}
\Phi_j(t,t') = \mu_j (t-t') + \phi_c(t)-\phi_c(t').
\end{equation}
Finally, the hybridizations $\Gamma_j$ were defined in Eq.~\eqref{hyb}.

While Eq.~\eqref{keld}, with the action in Eqs.~\eqref{sc11} and \eqref{sf11},
provides a general representation of the
interacting nonequilibrium Majorana transport problem, approximations are
necessary in order to obtain concrete analytical results.

\subsection{Semiclassical expansion}
\label{sec3b}

We here are mostly interested in the weak CB regime, where 
standard arguments\cite{dot} imply that
 a semiclassical approximation in the phase representation is appropriate.
In particular, fluctuations of $\phi_q$ around 
zero are small, i.e., we can expand the action in Eq.~\eqref{keld}
to quadratic order in $\phi_q$.  Keeping the full nonlinear $\phi_c$ dependence,
the matrix \eqref{antilamb} is thus approximated by 
$\check \Lambda_{j=L/R} \simeq \check \Lambda_{j}^{(0)} + 
\check \Lambda_{j}^{(1)} +\check \Lambda_{j}^{(2)}.$
Specifically, the matrices $\check \Lambda_{j}^{(m)}(t,t')$ of 
order $\phi_q^m$ are
\begin{widetext} 
\begin{eqnarray*}
\check \Lambda_{j}^{(0)} &=& i\Gamma_j
\left( \begin{array}{cc} 2F(t-t') \cos\Phi_j(t,t') & \delta_-(t-t')\\
-\delta_+(t-t') & 0\end{array}\right),\\
\check \Lambda_{j}^{(1)} &=& i\Gamma_j F(t-t') \sin\Phi_j(t,t')
\left( \begin{array}{cc} 0 & \phi_q(t') \\
-\phi_q(t) & 0\end{array}\right),\\
\check \Lambda_{j}^{(2)} &=& -i \frac{\Gamma_j}{4} 
F(t-t')\cos\Phi_j(t,t') \left( \begin{array}{cc} \phi^2_q(t)+
\phi_q^2(t') & 0 \\ 0 & -2\phi_q(t)\phi_q(t') \end{array}\right) 
+ \frac{\Gamma_j}{2} \phi_q^2(t) \delta(t-t') \check\tau_y.
\end{eqnarray*}
\end{widetext}
At this stage, it is convenient to introduce the 
\textit{interacting}\ Keldysh-Majorana Green's function (GF)
\begin{equation}\label{gj}
\check G_j(t,t') 
= \left(\check\tau_x i\partial_t+\check\Lambda_j^{(0)}\right)^{-1} .
\end{equation}
This has a triangular representation in (rotated) Keldysh space,  
$\check G_j = \left ( \begin{array}{cc} 0 & G_j^A \\ G_j^R & G_j^K 
\end{array} \right),$
with retarded/advanced GFs $G_j^{R/A}$ and Keldysh GF 
$G_j^K$.  Note that in this representation, 
the Majorana GF has a bosonic Keldysh structure,
see also Ref.~\onlinecite{golub}.
 According to Eq.~\eqref{gj},
$G^{R/A}_j$ obeys the noninteracting ($\phi_c=0$) equation of motion
and thus has the Fourier representation
$G_j^{R/A}(\epsilon) = 1/(\epsilon\pm i\Gamma_j)$. This
implies the spectral function 
\begin{equation}\label{spectral}
A_j(\epsilon) = - {\rm Im} \left[G_j^R(\epsilon)\right]= \frac{\Gamma_j}
{\epsilon^2+\Gamma_j^2} .
\end{equation} 
Only $G^K_j$ is affected by interactions encoded in $\phi_c$.

The action \eqref{sf11} now takes the form 
\begin{equation}\label{sm}
S_f = -i\sum_j \ln{\rm Pf}\left( \check G_j^{-1} \right)
+ S_f^{(1)}+S_f^{(2)}+ O(\phi_q^3),
\end{equation}
where the first ($\phi_q=0$) term vanishes as a result
of the unitary time evolution along the closed Keldysh contour. 
The first-order (in $\phi_q$) contribution is
\begin{equation}\label{sm1}
S_f^{(1)}= -\frac{i}{2}\sum_j {\rm Tr} \left( \check G_j \check\Lambda^{(1)}_j
\right) =  \int dt\ {\cal I} \phi_q, 
\end{equation}
where the trace ``Tr'' extends both over time and Keldysh space, and
${\cal I}$ denotes the total current flowing into the dot. Some algebra
yields
\begin{widetext}
\begin{equation}  \label{curr1}
{\cal I}(t) = \sum_j \Gamma_j\int dt' \ G_j^R(t-t') 
\left\{ F_{j,s}(t'-t)\cos\left[\phi_c(t')-\phi_c(t)\right]
+F_{j,a}(t'-t)\sin\left[\phi_c(t')-\phi_c(t)\right]\right\}
\end{equation}
\end{widetext}
with the time-symmetric (-antisymmetric) functions 
\begin{equation}\label{fsa}
F_{j,s}(t)=F(t)\sin(\mu_j t),\quad F_{j,a}(t)=F(t)\cos(\mu_j t).
\end{equation}
Finally, the second-order term in Eq.~\eqref{sm} reads
\begin{equation}\label{sm2}
S_f^{(2)}= -\frac{i}{2}\sum_j {\rm Tr}\left[
\check G_j\check\Lambda_j^{(2)}-\frac12 \left(\check G_j\check\Lambda_j^{(1)}
\right)^2 \right].
\end{equation}

\section{Weak Coulomb blockade limit}
\label{sec4}

\subsection{Langevin approach} \label{sec4a}

When the charging energy $E_c$ is sufficiently small, fluctuations of the
phase fields $\phi_{c,q}$ are small and allow for a quadratic expansion 
of the action $S_f$ in \textit{both}\ fields.\cite{dot}  This expansion
is detailed in Appendix \ref{appc} and allows to reformulate the 
functional integral (\ref{keld}) in terms of the
\textit{equivalent}\ semiclassical Langevin equation, 
\begin{equation}\label{langevin}
\frac{1}{2E_c} \ddot{\phi}_c(t)
 + \int^t dt' \ \eta(t-t') \dot \phi_c(t')= \xi(t).
\end{equation}
The \textit{damping kernel}\ $\eta(t-t')$ has the Fourier representation
\begin{widetext}
\begin{equation}\label{damp}
\eta(\omega) = \sum_{j=L/R} \frac{\Gamma_j}{\omega} \int\frac{d\epsilon}{2\pi}
\left( A_j(\epsilon+\omega/2) + A_j(\epsilon-\omega/2) \right)
\left [ F_{j,a}(\epsilon+\omega/2)-F_{j,a}(\epsilon -\omega/2) \right ].
\end{equation}
The Gaussian noise field $\xi(t)$ has zero mean value and the
correlation function $\langle \xi(t)\xi(t')\rangle= K(t-t')$, 
where the \textit{fluctuation kernel}\ has the Fourier representation
\begin{eqnarray} \label{kt}
K(\omega) &=& \sum_{j} \frac{\Gamma_j}{2}\int \frac{d\epsilon}{2\pi}
\left (A_j(\epsilon+\omega/2)+A_j(\epsilon-\omega/2)\right)
 \left[ 1- F_{j,a}(\epsilon+\omega/2) 
F_{j,a}(\epsilon-\omega/2)\right ] \\  \nonumber
&-& \sum_{j} \Gamma_j^2 \int \frac{d\epsilon}{2\pi} \ {\rm Re}
\left[ G_j^R(\epsilon+\omega/2) G^R_j(\epsilon-\omega/2)\right]
F_{j,s}(\epsilon+\omega/2) F_{j,s}(\epsilon-\omega/2).
\end{eqnarray}
\end{widetext}
In equilibrium ($\mu_{L,R}=0$),  $F_{j,s}=0$ and $F_{j,a}=F$,
and hence the fluctuation-dissipation relation 
\begin{equation}\label{fdt}
K_{\rm eq}(\omega) =\frac{\omega}{2}
 \coth\left(\frac{\omega}{2T}\right) \eta_{\rm eq}(\omega)
\end{equation}
is satisfied, providing an important consistency check.

In the zero-temperature limit, both $\eta(\omega)$ and $K(\omega)$
can be evaluated in closed form.  Specifically, we find  for the 
damping kernel
\begin{equation}\label{dampt0}
\eta_{T=0}^{}(\omega)= \sum_{j,\pm} \frac{\Gamma_j}{\pi\omega}  
\tan^{-1}\left(\frac{\omega\pm\mu_j}{\Gamma_j} \right),
\end{equation}
while the fluctuation kernel is
\begin{widetext}
\begin{equation} \label{ko} 
K_{T=0}^{}(\omega) = \frac{|\omega|\eta(\omega)}{2} +\sum_j \frac{\Gamma_j}{2\pi} 
\Theta(2|\mu_j|-|\omega|) \left[ 
\tan^{-1}\left(\frac{|\mu_j|}{\Gamma_j}\right) + 
\tan^{-1}\left(\frac{|\mu_j|-|\omega|}{\Gamma_j}\right)+
\frac{\Gamma_j}{|\omega|}\ln
\frac{(|\mu_j|-|\omega|)^2+\Gamma_j^2}{\mu_j^2
+\Gamma_j^2}\right],
\end{equation}
\end{widetext}
where $\Theta$ is the Heaviside function. 

The Langevin equation \eqref{langevin} allows for an intuitive 
interpretation of the charge dynamics in this interacting Majorana system in 
terms of an effective $RC$ circuit.  Its solution is used below in order 
to compute the current-voltage relation.

\subsection{Current-voltage relation}

Next we outline the calculation of the dc current, $I\equiv I_L=-I_R$. 
By using Eq.~\eqref{curr1} and the noise average 
$\langle \cdots\rangle_\xi=\int {\cal D}\xi \ (\cdots) e^{-\frac12
\xi K^{-1}\xi}$, we find the (indeed $t$-independent) result
\begin{eqnarray*}  
I_j &=&  \Gamma_j\int dt' \ G_j^R(t-t') \Bigl [
F_{j,s}(t'-t) \langle\cos[\bar\phi_c(t')-\bar\phi_c(t)]
\rangle_\xi \\ &+& F_{j,a}(t'-t)\langle \sin[\bar\phi_c(t')-\bar
\phi_c(t)]\rangle_\xi\Bigr].
\end{eqnarray*}
$F_{j,a/s}$ is given in Eq.~\eqref{fsa}, and
\begin{equation}\label{sol}
\bar \phi_c(t) = \int^t d\tau \ D^R(t-\tau)\xi(\tau)
\end{equation} 
solves Eq.~\eqref{langevin} for a given noise trajectory, where
$D^R(t-\tau)$ is the retarded solution of Eq.~\eqref{langevin}
with $\xi(t)\to \delta(t-\tau)$.  Performing the Gaussian noise average, 
we obtain 
\begin{equation}\label{curfin}
I_j = \Gamma_j \int dt' \ G_j^R(t-t') F_{j,s}(t'-t) e^{-J(t-t')},
\end{equation}
where the noise correlations are encoded in $J(t)=J(-t)$, similar to  the
standard ``$P(E)$'' theory of dynamical CB,\cite{cb}
\begin{equation}\label{pefunc}
J(t-t') = \frac12 \left\langle \left [ \bar\phi_c(t)-
\bar\phi_c(t') \right]^2 \right\rangle_\xi.
\end{equation}
Note that in order to obtain $J(t-t')$ one first has to solve 
the Langevin equation (\ref{langevin}) for a given noise trajectory,
and then average over all noise realizations.

Since $J(t-t')\ge 0$,  interactions always decrease the current
\eqref{curfin} with respect to the 
noninteracting solution, $I_j^{(0)}$. The latter follows by 
setting $J\to 0$ in Eq.~\eqref{curfin}.  In fact, after Fourier transformation, 
we get the known result\cite{law,bolech,nilsson,flensberg,golub}  
\begin{equation} \label{i0}
I_j^{(0)}= \Gamma_j\int \frac{d\epsilon}{2\pi}\ F(\epsilon-\mu_j) 
A_j(\epsilon)
\end{equation}
with the Lorentzian spectral function in Eq.~\eqref{spectral}.
In linear response ($\mu_j\to 0$) and taking the zero-temperature limit, we
recover resonant Andreev reflection with quantized conductance
$G_{j}= e \partial I_{j}/\partial \mu_{j}=2e^2/h$.

Equation (\ref{curfin}) for the current-voltage relation in 
the weak Coulomb blockade regime is one of the main results of this work.
We use this expression later on to quantify the effect of 
Coulomb interactions ($E_c\ne 0$) on the transport properties in this 
Majorana system.   

\subsection{Retardation effects}
\label{sec4b}

The Langevin equation \eqref{langevin} in general contains
memory effects because of the frequency-dependence of the damping
kernel.  These are most pronounced for $T=0$, where $\eta(\omega)$
takes the form in Eq.~\eqref{dampt0}.  For $\omega\to 0$, this gives 
the finite value 
\begin{equation}\label{eta0}
\eta_0 = \frac{2}{\pi} \sum_j \frac{1}{1+(\mu_j/\Gamma_j)^2},
\end{equation}
while for $|\omega|\to \infty$, we have $\eta(\omega)\simeq 
\Gamma/|\omega|$ with $\Gamma=\Gamma_L+\Gamma_R$.  
On low frequency scales, $|\omega|< 
{\rm min}\left(\sqrt{\mu_j^2+\Gamma_j^2}\right)$, 
the kernel $\eta(\omega)$ can be approximated by $\eta(\omega)\simeq \eta_0$.
This approximation corresponds to the absence of retardation in 
the damping kernel $\eta(t)$. This approximation
works even better for finite $T$, where
\begin{equation}\label{eta1}
\eta_0 = \sum_j \frac{\Gamma_j}{T}
 \int\frac{d\epsilon}{2\pi}\frac{1}{\cosh^2\left( \frac{\epsilon}{2T}\right) } 
\frac{\Gamma_j}{(\epsilon-\mu_j)^2+\Gamma_j^2}
\end{equation}
yields a damping suppression, 
$\eta_0\simeq \Gamma/(2T)$, in the high-$T$ limit. 

We therefore employ the simplification $\eta(\omega)=\eta_0$ from now on,
which is appropriate to describe low-energy transport.
This implies from Eq.~\eqref{langevin} that we now have 
a Markovian Langevin equation,
\begin{equation}\label{lang2}
 \ddot{\phi}_c(t)+ \Omega \  \dot{\phi}_c(t) = 2E_c \ \xi(t),
\end{equation}
and $D^R(t)$ in Eq.~\eqref{sol} simplifies to
\begin{equation}\label{dr2}
D^R(t) = \frac{1}{\eta_0} \left(1-e^{-\Omega t}\right) \Theta(t).
\end{equation}
The inverse ``$RC$ time'' in Eq.~\eqref{lang2} is given by
\begin{equation}\label{omega}
\Omega= (RC)^{-1} = \eta_0 E_c.
\end{equation}
Since $\eta_0\le 4/\pi$, see Eq.~\eqref{eta0}, 
and using $E_c=e^2/(2C)$, the effective resistance $R$
in Eq.~\eqref{omega} always fulfills $R\ge h/(4e^2)$.  Note that 
this corresponds to a parallel resistance as seen by the dot.  

However, at the same time it is essential to keep the 
full frequency-dependence in the fluctuation kernel $K(\omega)$ 
when evaluating $J(t)$  in Eq.~\eqref{pefunc}.  
Using Eq.~\eqref{dr2}, some algebra yields 
\begin{equation}\label{j2}
J(t) =\frac{1}{\pi\eta_0^2} \int_0^\infty d\omega \
K(\omega) \frac{1-\cos(\omega t)}{\omega^2 (1+\omega^2/\Omega^2)},
\end{equation}
where the current $I=I_L$ follows from Eq.~\eqref{curfin}. 

\subsection{Zero-bias anomaly}

We now present results for the $IV$ relation in  
the zero-temperature limit, where MBS effects
are most pronounced.  We adopt the Markovian approximation
described in Sec.~\ref{sec4b} and study a symmetric 
system, where $\Gamma_L=\Gamma_R=\Gamma/2$ and $\mu_L=-\mu_R=eV/2$.
As relevant transport quantity, we define the 
dimensionless nonlinear conductance 
\begin{equation}\label{gv}
g(V) = \frac{I(V)}{e^2 V/h}.
\end{equation}
For $E_c=0$, the exact solution [Eq.~\eqref{i0}] yields
\begin{equation}\label{g00}
g^{(0)}(V) = \frac{\Gamma}{eV} \tan^{-1} \left(\frac{eV}{\Gamma}\right).
\end{equation}
Note that the unitary limit (resonant Andreev reflection)
corresponds to $g=1$, which is reached for $V\to 0$ in Eq.~\eqref{g00}.

\begin{figure}
\centering
\includegraphics[width=8cm]{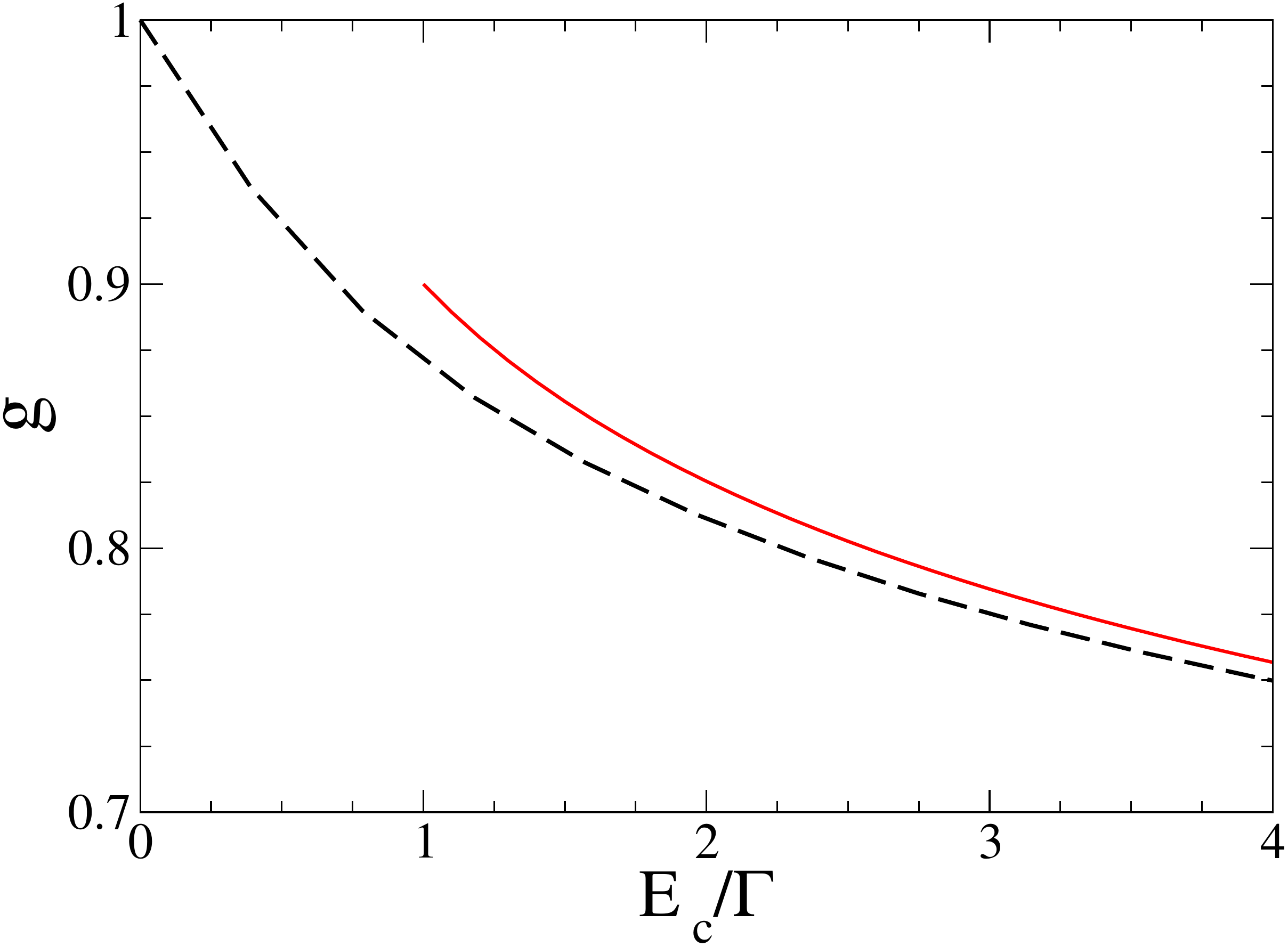}
\caption{\label{fig2} (Color online)
Dimensionless linear conductance, $g=g(V\to 0)$, vs $E_c/ \Gamma$ 
in the weak CB regime. The black dashed curve has been obtained from 
numerical analysis of Eq.~\eqref{curfin}, with $J(t)$ in 
Eq.~\eqref{j2} and $K(\omega)$ in Eq.~\eqref{ko}. The red solid curve
gives the analytical result in Eq.~\eqref{g1} valid for $E_c\agt \Gamma$.
}
\end{figure}

\begin{figure}
\centering
\includegraphics[width=8cm]{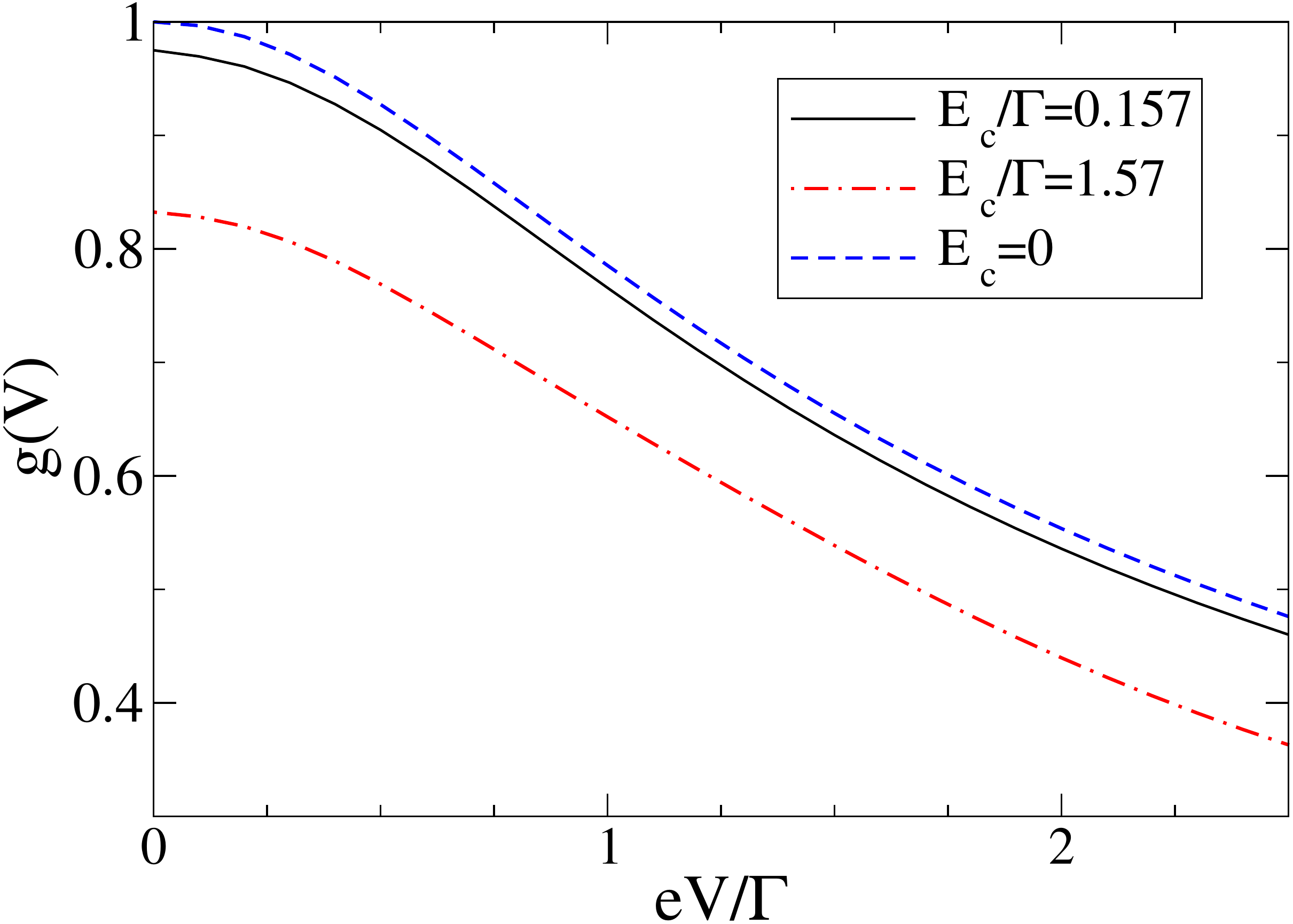}
\caption{\label{fig3} (Color online)
Dimensionless nonlinear conductance $g(V)$ vs $eV/\Gamma$ for
several values of the charging energy. 
Results for $E_c>0$ were obtained by numerical integration, the
result for $E_c=0$ is exact.  }
\end{figure}

We now write $J(t)=J_0+J_1$, where 
\begin{equation}
J_0(t) = \frac{1}{2\pi\eta_0}\int_0^\infty \frac{d\omega }{\omega}
\frac{1-\cos(\omega t)}{1+\omega^2/\Omega^2}
\end{equation}
comes from the first term ($\propto \eta(\omega)=\eta_0$) in Eq.~\eqref{ko}.
$J_1$ is a pure nonequilibrium term: the relevant kernel contribution
scales as $\sim V^3$.  Let us then first consider the 
\textit{linear response regime}, where $J_1$ can be discarded.
To logarithmic accuracy, $J_0(t)\simeq 
(2\pi\eta_0)^{-1}\ln(\Omega t)$ for $\Omega t> 1$,
where $\eta_0=4/\pi$ within linear response, see Eq.~\eqref{eta0}. 
Equation \eqref{curfin} then yields the dimensionless linear conductance 
in analytical form,
\begin{equation}\label{g1}
g(V\to 0)= c_0 (E_c/\Gamma)^{-1/8} , \quad c_0\approx 0.96.
\end{equation}
This result holds\cite{foot2} for $\Gamma\alt E_c$.  
As a function of $E_c/\Gamma$, Eq.~\eqref{g1} reveals a \textit{power-law 
suppression}\ of the linear conductance with the universal exponent $1/8$,
reminiscent of a zero-bias anomaly.
For arbitrary $E_c/\Gamma$, the current given by Eq.~\eqref{curfin}
can be evaluated numerically.
The resulting linear conductance is shown in Fig.~\ref{fig2}, and the
analytical result \eqref{g1} is quite accurate 
in describing the interaction ($E_c\ne 0$) induced suppression 
of the current for $E_c\agt \Gamma$.

Next we turn to the nonlinear conductance $g(V)$ in Eq.~\eqref{gv}.
Numerical results for several values of $E_c/\Gamma$ are shown
in Fig.~\ref{fig3}, where the exact result [Eq.~\eqref{g00}] 
for $E_c=0$ is also depicted.  We find  a clear suppression
of $g(V)$ when the applied bias voltage increases.  This
suppression becomes stronger with increasing $E_c$. In fact, we observe
that the interaction induced suppression of the current is more pronounced for 
small voltage.  We therefore interpret our results as a zero-bias 
anomaly caused by the (weak) Coulomb blockade of Majorana fermion 
induced transport.

\section{Concluding remarks}
\label{sec5}

In this work, we have presented a general theory of transport
through a mesoscopic superconductor containing a pair of
MBSs.  We have studied the regime of low-energy transport, 
where metallic leads are attached and all relevant energy scales are small
compared to the superconducting gap.  The case of weak
Coulomb blockade, where the charging energy $E_c$ does not significantly
exceed the typical hybridization scale $\Gamma$, has been investigated
in some detail, and we found clear signatures for an interaction ($E_c\ne 0$)
induced suppression of both the linear and the nonlinear conductance.

We have introduced a formulation to analyze interaction
effects in electronic transport through MBSs which does not rely
on the Hilbert space constraints of previous approaches and thus is easier
to handle in general.  We are presently extending this formulation to
study several other situations of current interest. 
In particular, relevant questions
include the full counting statistics and the nonequilibrium dephasing
in multi-terminal transport geometries containing MBSs, and the
Josephson effect with superconducting leads.  Furthermore, the
strong Coulomb blockade regime also allows for the detailed 
study of the full Keldysh functional integral \eqref{keld} 
in terms of a rate equation approach. 

To conclude, we are confident that the characteristic features 
reported here, i.e., the suppression of the linear conductance with 
increasing $E_c/\Gamma$ and of the nonlinear 
conductance with $V/\Gamma$ in the limit of weak Coulomb blockade, 
can be observed experimentally once MBSs have been detected.
The setup studied here may allow to find signatures of 
the elusive Majorana fermions in transport properties, including
phenomena such as resonant Andreev reflection, teleportation, and 
the universal power-law suppression of the linear conductance,  
$g\sim (E_c/\Gamma)^{-1/8}$, for intermediate values of the ratio
$E_c/\Gamma$.

\acknowledgments

This work was supported by the SFB TR 12 of the DFG 
and by the Spanish MICINN under contract FIS2008-04209.

\appendix
\section{Derivation of the tunnel Hamiltonian}
\label{appa}

In order to keep the paper self-contained, we here provide a
derivation of the tunnel Hamiltonian (\ref{ht1}). 
Following Flensberg,\cite{flensberg} the field operator $\Psi_\sigma(x)$
for electrons with spin projection $\sigma=\uparrow,\downarrow$
in a TS wire of length $L$ (with $0<x<L$) 
defines the Majorana fermion operators
\[
\gamma_{j=L/R} = \sum_\sigma\int_0^L dx \ \left[ f^*_{j\sigma}(x)\Psi_\sigma(x)
+ f_{j\sigma}(x) \Psi_\sigma^\dagger(x) \right]
\]
with the MBS wavefunction $f_{j\sigma}(x)$. Projecting the full Nambu
spinor onto the MBS subspace, one obtains\cite{flensberg}
\[
\Psi_\sigma(x)\to \sum_{j} f_{j\sigma}(x) \ \gamma_j .
\]
The tunnel Hamiltonian now reads 
\begin{eqnarray*}
H_t &=& \sum_{jk\sigma} \int_0^L 
dx \  t^*_{j}(x) \ C^\dagger_{jk\sigma} \Psi_\sigma(x) + {\rm h.c.}
\\ &=& \sum_{jk\sigma} V_{j\sigma}^* C^\dagger_{jk\sigma}\gamma_j +
{\rm h.c.},
\end{eqnarray*}
where $C^\dagger_{jk\sigma}$ creates an electron in lead $j=L/R$ with
momentum $k$ and spin projection $\sigma$.  Here
$t_j(x)$ is assumed $k$-independent for simplicity, which defines
the tunnel matrix elements $V_{j\sigma}= \int dx \ t_j(x) f^*_{j\sigma}(x)$.
Note that one can always form suitable
linear combinations of $C_{jk,\uparrow}$ and $C_{jk,\downarrow}$ 
to form spinless lead fermions $c_{jk}$ coupled to the MBSs,
\[
\sum_{\sigma}  V^*_{j\sigma} C^\dagger_{jk\sigma} \to \lambda_j 
c_{jk}^\dagger.  
\]
The other (orthogonal) linear combination then decouples from the problem.
Finally, inserting the auxiliary fermion representation (\ref{aux})
and taking into account charge conservation, we arrive at the tunnel
Hamiltonian quoted in Eq.~\eqref{ht1}.

\section{On the Keldysh functional integral}
\label{appb}

Here we provide a detailed derivation of the AES action in Sec.~\ref{sec3a}.
We start by constructing the Lagrangian $L_c$ for the isolated dot in terms
of $\phi(t)$ and the Grassmann variable $d(t)$,
plus the corresponding ``velocities'' $\dot \phi=\frac12 
\partial H_c/\partial N$
and $\dot d$. Noting that for Grassmann variables, $i\bar d$ is canonically
dual to $d$, we have
\[
L_c= \frac{\dot \phi^2}{4 E_c} + n_0 \dot\phi + \bar d 
(i\partial_t-\dot\phi) d.
\]
Adding the tunnel contribution and performing the gauge transformation
$d\to e^{-i\phi} d$, we obtain
\begin{eqnarray*}
L_c+L_t &=& 
\frac{\dot \phi^2}{4 E_c} + n_0 \dot\phi + \frac{i}{2}\sum_j\gamma_j\dot 
\gamma_j \\
&-& \sum_{jk} \left( \lambda_j \bar \psi_{jk} e^{-i\phi} \gamma_j + {\rm h.c.}
\right).
\end{eqnarray*}
Here we switched back from the auxiliary $d$ fermion to the Majorana
field $\gamma_j$. Note that up to a full time derivative,
 $i\bar d \dot d\to (i/2)\sum_j \gamma_j \dot \gamma_j$.
The Grassmann fields $(\psi_{jk},\bar\psi_{jk})$ correspond to the 
lead fermion operators $(c^{}_{jk},c^\dagger_{jk})$.
Using the Keldysh formulation, we double all fields according
to the forward and backward branch of the Keldysh time contour, i.e.,
$\gamma_j(t)\to (\gamma_{j,+}(t),\gamma_{j,-}(t))^T$ and so on.
It is also convenient to gauge out the chemical potentials $\mu_j$ in the
leads, $\psi_{jk,\pm}(t)\to e^{i\mu_j t} \psi_{jk,\pm}(t)$. 
We now use the Keldysh matrix notation 
$\check \phi(t)=\phi_c+\check\tau_z \phi_q/2$,
see Eq.~\eqref{phicq}. The complete Keldysh action, $S=S_\gamma+S_c+S_t+S_l$, 
contains $S_c$ in Eq.~\eqref{sc11} and the pieces 
\begin{eqnarray*} 
S_\gamma &=& \frac{i}{2} \sum_j \int dt \ 
\gamma_j \check \tau_z \dot\gamma_j,\\
S_t &=& - \sum_{jk} \lambda_j \int dt \ \bar\psi_{jk}\check\tau_z
e^{-i[\mu_j t+\check\phi(t)]} \gamma_j + {\rm h.c.}, \\
S_l &=&\sum_{jk} \int dt \ \bar\psi_{jk}\check\tau_z
(i\partial_t-\epsilon_{jk})\psi_{jk},
\end{eqnarray*}
where $\epsilon_{jk}$ refers to the dispersion relation in lead $j$.

The current $I_j$ flowing from lead $j$ into the dot follows from
the Heisenberg equation of motion. The currents obey the relation
$(I_L+I_R)(t) =  \delta S_t/\delta \phi_q(t)$. 
Current conservation implies $\langle I_L\rangle =- \langle I_R\rangle$, 
which fixes the mean value $\langle\dot \phi_c\rangle$ to
the chemical potential $\mu_s$ of the superconducting dot. The
latter has to be determined self-consistently from current conservation.
We then redefine $\phi_c\to \mu_s t +\phi_c(t)$, i.e.,
$\phi_c(t)$ now refers to fluctuations around the mean-field
value ($\mu_s t$) of the classical phase variable. 
The ensuing changes in $S_c$ and $S_t$ can be absorbed in a
renormalization of $n_0$ and $\mu_j$. In particular, denoting their
bare values by $\bar n_0$ and $\bar \mu_j$, respectively, we
find $n_0=\bar n_0+\mu_s/(2E_c)$ and $\mu_j=\bar \mu_j+\mu_s$.

The next step is to integrate out the Grassmann fields 
$(\psi_{jk},\bar\psi_{jk})$. This is a standard step\cite{naz,altl,dot,altland}
and leads to the effective Majorana action $S_{\rm eff}$ 
replacing $S_l+S_t+S_\gamma$.
Before turning to the result, we perform the usual rotation in Keldysh space
in order to have triangular Green's function (GF) representations,
using the unitary matrix\cite{naz}
$\check L= \frac{1}{\sqrt{2}} \left(\begin{array}{cc} 1 & -1\\
1 & 1\end{array}\right).$
The rotated Majorana fields are $\tilde\gamma_j = \check L \gamma_j$. 
Evaluating the resulting momentum integrals for the leads
in wide-band approximation, we find the effective Majorana action
\begin{widetext}
\begin{eqnarray*}
S_{\rm eff}  &=& \frac12 \sum_j \int dt dt'  \tilde \gamma_j^T(t) 
\left[ \check \tau_x i\partial_t  \delta(t-t')+
\check Q_j(t,t')\right] \tilde\gamma_j(t') \\
\check Q_j(t,t') & = & i\Gamma_j e^{i\check\tau_x \phi_q(t)/2}
\left(\begin{array}{cc} \delta_-(t-t') & 2F(t-t')e^{i\Phi_j(t,t')}\\
0 & -\delta_+(t-t') \end{array}\right) \check\tau_x 
e^{-i\check\tau_x \phi_q(t')/2}.
\end{eqnarray*}
\end{widetext}
The hybridizations $\Gamma_j$ are defined in Eq.~\eqref{hyb}
with lead density of states $\nu_j=\sum_k \delta(\epsilon_{jk})$,
$F(t)$ has the Fourier transform \eqref{dist},
and the phase function $\Phi_j$ is specified in Eq.~\eqref{Phi}.

Now we are ready to also integrate out the Majorana fields. This
finally yields Eq.~\eqref{keld} where
\[
\check \Lambda_j(t,t')= \frac12 \left[ \check Q_j(t,t') -
\check Q^T_j(t',t) \right]
\]
is the antisymmetric part of $\check Q_j$.

\section{Derivation of Langevin equation}
\label{appc}

Here we provide some details concerning the derivation of the Langevin equation 
\eqref{langevin} in Sec.~\ref{sec4a}. 
We first analyze the action piece $S_f^{(1)}=\int dt\ {\cal I} \phi_q$.
As mentioned in App.~\ref{appb}, the chemical potential $\mu_s$ of the 
dot has to be chosen self-consistently in order to fulfill 
current conservation.
By construction, we must therefore have the average value ${\cal I}=0$
when $\dot \phi_c=0$. Using Eq.~\eqref{curr1}, expanding to first order 
in $\phi_c$, and performing a partial integration, we find
\[
S_f^{(1)} =-\int dt\ \phi_q(t) \int^t dt' \ \eta(t-t') \dot \phi_c(t'),
\]
where $\eta(t-t')$ is determined by Eq.~\eqref{damp}. 
Next we note that $S_f^{(2)}$, which is already
of order $\phi_q^2$, can be evaluated by replacing the 
interacting Keldysh GF $G^K_j$ with its noninteracting ($\phi_c=0$) 
version.  After some algebra, we find from Eq.~\eqref{sm2} 
\[
S_f^{(2)} = \frac{i}{2}\int dt dt' \ \phi_q(t) K(t-t') \phi_q(t')
\]
with $K(t-t')$ determined by Eq.~\eqref{kt}.  The
Fourier-transformed kernel $K(\omega)$ directly describes finite-frequency
current noise correlations for $E_c=0$, a question that has
been studied for $\omega\to 0$ in Ref.~\onlinecite{golub}.
Finally, we complete the derivation by Hubbard-Stratonovich 
transformation of the $\phi_q^2$ term
in the generating functional [Eq.~\eqref{keld}],
\[
e^{iS_f^{(2)}}= \int {\cal D} \xi \ e^{-\frac12\int dt dt' \ \xi(t) K^{-1}(t-t')
\xi(t') + i \int dt\ \phi_q \xi}.
\]
Functional integration over $\phi_q$, which appears only linearly
in the action, yields Eq.~\eqref{langevin}, and the Gaussian noise field 
$\xi(t)$ indeed has the correlation function $K(t-t')$ in Eq.~\eqref{kt}.


\begin{thebibliography}{99}
\bibitem{foot2} 
The weak CB regime is eventually left for $\Gamma\ll E_c$, 
where Eq.~\eqref{g1} ceases to be valid.
\end{thebibliography}
\end{document}